\newcommand{\be}{\begin{equation}}
\newcommand{\ee}{\end{equation}}

\documentclass[aps,pre,twocolumn]{revtex4}

\usepackage{amssymb}
\usepackage{graphicx}
\usepackage{dcolumn}
\usepackage{bm}
\usepackage{amsmath}

\begin{document}
\title{Optical guiding of absorbing nanoclusters in air}

\author{Vladlen G. Shvedov$^{1,2,3}$, Anton S. Desyatnikov$^1$, Andrei V. Rode$^3$, Wieslaw Z. Krolikowski$^3$ \& Yuri S. Kivshar$^1$}

\affiliation{$^1$Nonlinear Physics Center, Research School of Physics and Engineering,\\Australian National University, Canberra ACT 0200, Australia\\
$^2$Department of Physics, Taurida National University, Simferopol 95007 Crimea, Ukraine\\
$^3$Laser Physics Center, Research School of Physics and Engineering, Australian National University, Canberra ACT 0200, Australia}

\maketitle

{\bf
Two effects of the light-matter interaction are widely used for micromanipulation of particles with laser beams~\cite{tweezers08}: radiation pressure~\cite{AshPRL70}, originating from a direct transfer of momentum from photons, and electric dipole (gradient) force acting on polarized particles~\cite{AshOL86}. In gaseous media, however, the heating of absorbing particles by light leads to much stronger radiometric forces~\cite{book2002}  considered until now only as an obstacle which prevents a stable trapping~\cite{AshSCI80}. In contrast, here we show that photophoretic force~\cite{PZ_17_352,Preining_66} can be specifically tailored to trap and manipulate absorbing particles in open air. We demonstrate experimentally the optical guiding of clusters of carbon nanoparticles~\cite{14,15} in the form of nanofoam fragments of arbitrary shape, with the size in the range 0.1-10 micrometers, and for laser powers lower than one milli Watt. The optical trap is created by two counter-propagating ``doughnut'' vortex beams~\cite{Berry,Soskin}, and it allows simultaneous trapping of several particles as well as their stable positioning and controlled guiding along the optical axis. Only a small fraction of operating power is actually absorbed because the particles are trapped in the region of vanishing intensity at the vortex core. Thus the alteration of physical and chemical properties of airborne particles is minimal in the trap, this feature is important for experiments with aerosols~\cite{EPJE_04_287}. Furthermore, the non-contact and remote optical trapping of absorbing aerosol particles can be applied to detect, monitor, and possibly reduce the exposure to engineered nanomaterials in air~\cite{OberdorPatFibTox2005}, a necessary tool~\cite{MaynardNat2006} for the studies of impact of nanotechnology on health~\cite{NelSci2006} and environment~\cite{ColvinNatBio2003}. The photophoretic trapping can be also employed to simulate, on laboratory scales, the processes studied in atmospheric~\cite{JGR_67_455,PRL_06_134301} and planetary~\cite{AA_07_977, AA_07_L9} sciences.}


When a photon is absorbed by a small particle its momentum contributes towards radiation pressure while its energy dissipates in heat. The latter leads to thermal forces~\cite{book2002} deliberately avoided in optical tweezers~\cite{tweezers08}: an optical trap utilizing a gradient force produced by a single strongly focused laser beam~\cite{AshOL86}. The applications of optical tweezers range from trapping of colloidal particles~\cite{AQC_98_469,tweezers03} and living cells~\cite{bio} to manipulation of single molecules~\cite{molecule} and atoms~\cite{chu}. Yet these applications are limited by the condition of vanishing radiometric forces and exclude, for example, trapping of strongly absorbing particles in air~\cite{AQC_98_469,air}. Indeed, if the surface of an aerosol particle is nonuniformly heated by an incident light, the gas molecules rebound off the surface with different velocities creating an integrated force on the particle, this effect was discovered by Ehrenhaft~\cite{PZ_17_352} and termed {\em photophoresis} (PP)~\cite{book2002,Preining_66}. A rough comparison~\cite{APL_82_455} of the radiation pressure force, $F_a=P/c$, exerted by a beam with power $P$, and the PP force, $F_{\rm pp}=P/3v$, for particles with zero thermal conductivity~\cite{JGR_67_455}, shows that for air at room temperature the later dominates by several orders of magnitude, $F_{\rm pp}/F_a=c/3v\simeq 6\times10^5$, here $c$ is the speed of light and $v$ is the gas molecular velocity. More elaborate calculations~\cite{BeresnevPhFl1993}, which we provide below in Methods and Supplementary Notes, show that the PP force acting on carbon nanoclusters in air is four orders of magnitude larger than $F_a$. The radiometric trapping of particles against gravitation, or optical levitation~\cite{APL_82_455}, was demonstrated~\cite{JCS_64_50}  several years prior to the first experiments on trapping with radiation pressure~\cite{AshPRL70}, but stable all-optical trapping of particles utilizing strong PP forces was not realized yet. At the same time, opaque particles are of main concern for environmental protection, e.g., the familiar air pollutants, such as car exhaust, as well as novel fabricated nanomaterials, such as carbon nanotubes and metallic nanoparticles.

The major difficulty in utilizing PP force for particle trapping is that a finite time of thermal relaxation within the particle is sufficient for its stochastic motion along complex trajectories~\cite{Preining_66}. Nevertheless, it has been experimentally observed~\cite{AQC_98_469,APL_82_455} that using a vortex beam with the ring-shaped transverse intensity profile~\cite{Berry,Soskin} leads to a strong two-dimensional confinement of absorbing particles on the ``dark'' optical axis. The key step forward that we make here for realizing a fully three-dimensional trapping is the implementation of the dual-beam scheme~\cite{AshPRL70} but with co-rotating counter-propagating vortex beams~\cite{Shvedov}, see Fig.~\ref{fig1}({\bf a}). The longitudinal on-axis confinement is achieved by a balance of PP forces induced by two beams on the opposite sides of a particle while the transverse confinement by the bright intensity ring compensates for gravity in the horizontal scheme. The details of the experimental setup can be found in Methods Summary and Supplementary Methods.

\begin{figure}
\centering\includegraphics[width=8.5cm]{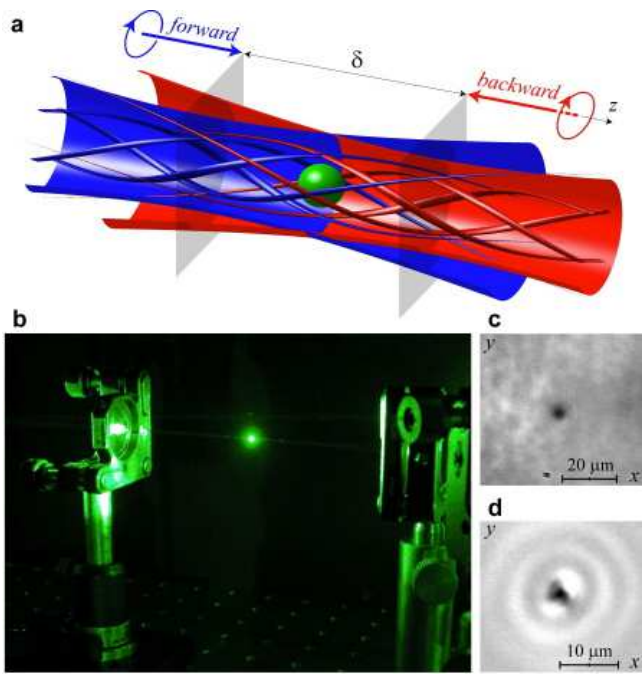}
\caption{{\bf Photophoretic trap.} {\bf a}, Schematic of an optical trap with counter-propagating and co-rotating vortex beams shown by surfaces at their tube-like intensity maxima. The focal (gray) planes of the froward (blue) and backward (red) beams are separated by the distance $\delta$, for equal powers of two beams the trapping position is in the middle between two planes. Particle (green sphere) is subject to illumination from both sides, the geometry of the laser power flow (arrows) is shown with the stream-tubes, the varying width of tubes is proportional to the modulus of the Poynting vector. {\bf b}, {\bf d}, The shade cast by a trapped particle as seen on the optical axis ({\bf b}) on white-light background and ({\bf d}) with superimposed vortex beam. {\bf c}, The side view of the setup with a particle trapped in air. A halo of the scattered light makes particle visible to a naked eye.}
\label{fig1}
\end{figure}

To realize the PP trapping we use clusters of agglomerated carbon nanoparticles~\cite{14,15} produced with high-repetition-rate laser ablation technique as described in Methods. The nanoclusters scatter sufficient amount of light to be visible by a naked eye, see Fig.~\ref{fig1}({\bf c}). We also use an additional white-light source to monitor the transverse dynamics of trapped particles, as seen in Figs.~\ref{fig1}({\bf b, d}) and corresponding Movies 1 and 2. The PP trapping is sufficiently robust to stabilize particle in open air for several minutes, as in Fig.~\ref{fig1}({\bf c}), and we use a glass cell with an open top to neutralize air draughts; once captured in the cell the particle remains trapped for many hours even when the operating power is reduced below one milli Watt. Additional challenge of trapping aerosols~\cite{air}, in contrast to colloids~\cite{tweezers08,AQC_98_469}, is that the trapping of freely moving particles is passive rather than active, i.e., we tap the cell containing powder of fragmented nanofoam so that small amount is released in air around the PP trap, and we wait till one or several particles remain captured stationary. In the transient regime, we observe fascinating scattering of many particles drawn towards the trapping region and competing for a stable position. Thus we have no control over the size or the shape of trapped particles.

The average size of individual nanoparticle range from 4~nm to 8~nm as seen in the inset in Fig.~\ref{fig2}({\bf a}). Figure~\ref{fig2} also shows the Transmission Electron Microscope (TEM) and Scanning Electron Microscope (SEM) images of nanoparticles ({\bf a}) near and ({\bf b}) far away from the ablated graphite target, the later nanofoam is used in our experiments as a source of aerosol particles. The SEM images of nanoclusters collected directly from the trap are presented in Figs.~\ref{fig2}({\bf c, d}). Among many collected particles of different shapes and sizes we choose to show one of the smallest in ({\bf c}), with the linear size of the order of 100~nm, and one of the largest in ({\bf d}), spanning for over 10~$\mu$m. Particles smaller than 100~nm remain invisible in our scheme, whether trapped or not, and the lower limit for the size of trapped particles is unknown. A simple theoretical model of the PP trapping which we suggest and describe in Supplementary Notes II shows that the trapping efficiency  decays rapidly  for particles larger than the vortex ring diameter $w$; in our experiment this value was of the order of $10\,\mu$m which agrees well with the size of the largest trapped particle.

\begin{figure}
\centering\includegraphics[width=8.5cm]{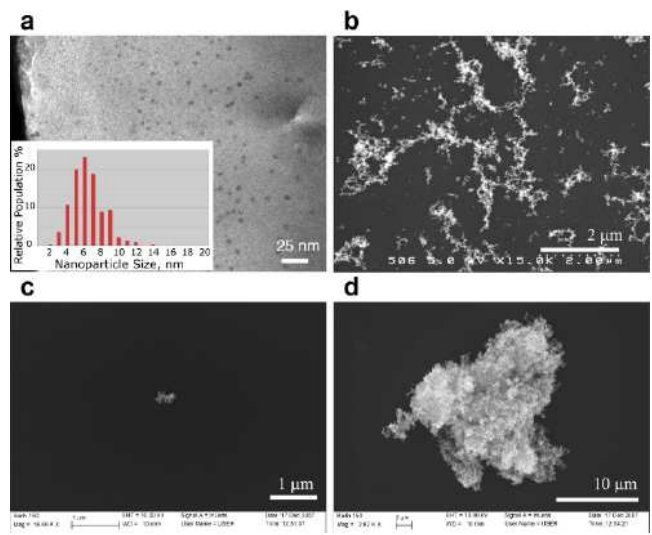}
\caption{{\bf Electron micrographs of carbon nanoclusters produced by laser ablation and collected from the optical trap.} {\bf a}, TEM micrograph of single nanoparticles collected on a holey carbon TEM grid in the laser ablation chamber at about 10~mm from the ablated graphite target. The inset shows the nanoparticle size distribution with the maximum at 6~nm. {\bf b}, SEM image of the nanoparticle aggregates deposited on a silicon substrate approximately 200~mm from the laser ablation area. {\bf c}, {\bf d}, SEM images of samples of carbon nanoclusters, composite of aggregated nanoparticles, collected from the photophoretic trap.}
\label{fig2}
\end{figure}

For optical guiding we modify the dual-beam trap~\cite{AshPRL70} and include an element of optical control over the axial position $Z$ of the particle. It consists of a half-wave plate and polarizing beam-splitter cube with low extinction ratio, approximately 1:13, see Supplementary Methods for the details. Varying the tilt angle $\theta$ of the half-wave plate we are able to change the power ratio $\varepsilon(\theta)=P_f/P_b$ of the forward ($P_f$) and backward ($P_b$) vortex components. Imbalance of the powers illuminating particle from both sides shifts the trapping position towards a weaker beam, this shift is limited by the cube extinction ratio. In Supplementary Notes we provide a theoretical model which predicts the location $Z$ of the trap, in a good agreement with experiment, as a function of the tilt angle $\theta$, the size of the spherical particle $a$, vortex ring radius $w$, and the distance separating focal planes of two beams $\delta$. In particular, we show that for $a < w/2$ the stationary position $Z$ does not depend on the particle radius $a$, and it is determined only by the separation $\delta$ and the power ratio $\varepsilon(\theta)$.

Experimentally we record the images of the radiation scattered by trapped particles in the direction perpendicular to the optical axis, the results are presented in Fig.~\ref{fig3} and Movie 3. First, we perform a static guidance when the particle is trapped stationary for each tilt $\theta$ changed with the step of $2^\circ$. The focal planes are separated by $\delta=2.0\pm 0.2\,\mu$m (s.d.e.), and we could pinpoint stably the particle on the beam axis anywhere within the distance of about 1~mm, see Fig.~\ref{fig3}({\bf a}). It is seen that the position is not a strictly periodic function of $\theta$, i.e., the particle does not return to the same position for a given value of $\theta$. For a better comparison, we perform similar experiments but with the maximum magnification of our imaging system, the results are presented in Figs.~\ref{fig3}({\bf b, c}). The half-wave plate is turned about one of the extrema indicated in Fig.~\ref{fig3}({\bf a}) by a dashed rectangle. The uncertainty in the trapping position versus $\theta$ reaches values $\sim 200\,\mu$m, greatly exceeding the uncertainty for the stationary particle, the vertical bars in Figs.~\ref{fig3}({\bf a, b}) indicate the recorded spot size $\sim 10\div20\,\mu$m. We rule out instability of trapping because we never observed a trapped particle to change its position spontaneously. The most plausible explanation for the observed hysteresis-type behavior is the presence of multiple traps produced by rather large distortion in the shape of vortex beams as well as misalignment of their axes. Nevertheless, it is always possible to position the particle with the precision limited only by the resolution of our imaging system by appropriately adjusting $\varepsilon(\theta)$.

\begin{figure}
\centering\includegraphics[width=8.5cm]{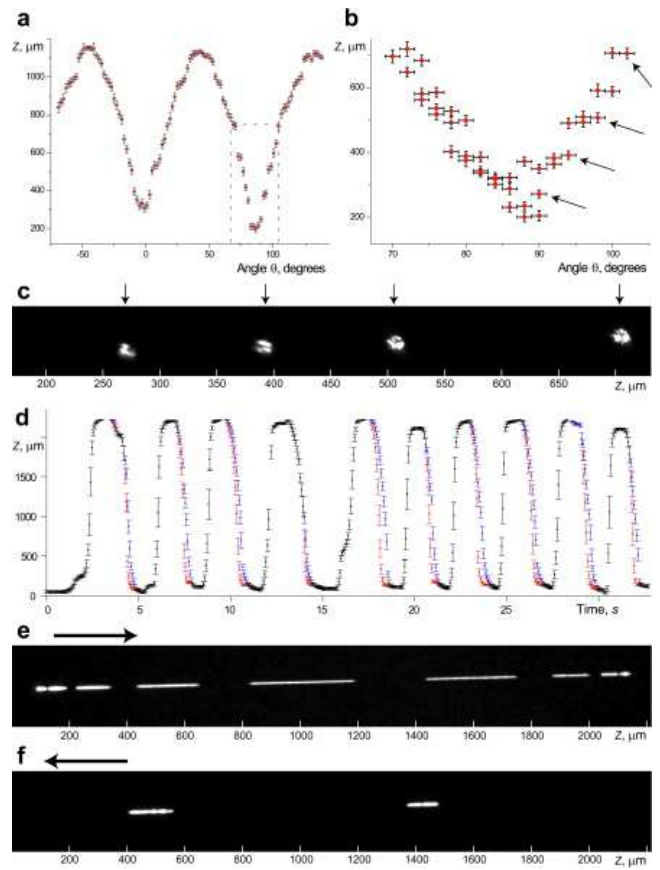}
\caption{{\bf Guiding of particles in air.} {\bf a-c}, Static guiding of particles: the position $Z$ of a trapped particle measured as a function of the polarizer angle $\theta$ in ({\bf a}) and with higher magnification in ({\bf b}). Vertical bars measure the spot size of the recorded particle images such as those superimposed in ({\bf c}); corresponding data points are marked in ({\bf b}) with arrows. {\bf d-f}, Dynamic guiding of particles: the positions of two particles simultaneously bouncing between two extremum points of the trap versus time are shown in ({\bf d}); black bars correspond to the on-the-fly tracks overlapping for both particles, such as the nine consecutive tracks superimposed in {\bf e}. Blue and red bars in ({\bf d}) measure the two tracks of separated particles, such as in the snapshot at the time $t=23$~s in ({\bf f}). Arrows in ({\bf e}) and ({\bf f}) indicate directions of particle propagation.}
\label{fig3}
\end{figure}

To test the dynamic stability of the PP trapping we performed hand-on experiments of continuous movement of particles recording on-the-fly tracks with exposition 40~ms. The half-wave plate was rotated changing $\varepsilon(\theta)$ periodically so that particle(s) were bouncing back and force over the distance exceeding 2~mm many times and reaching velocities up to 1~cm/s, here the inter-focal distance was $\delta=7.4\pm0.4$~mm (s.d.e.). The results are presented in Movie 4 and Figs.~\ref{fig3}({\bf d-f}). The position $Z$ versus real time in ({\bf d}) could be determined only with large uncertainty as the length of the tracks (vertical bars) recorded with finite exposition time. In this particular experiment there were two particles trapped simultaneously but they visibly separated only when moving in one direction, from the right to the left, as shown in the snapshot in ({\bf f}). At any moment of this experiment the rotation could be stopped and the particles stop and became indistinguishable. Similarly impossible is to distinguish two particles when their tracks overlap on the flight from the left to the right in ({\bf e}).
Asymmetry in our scheme, $0.093\le\varepsilon\le15.509$, is attributed to loses in one of the interferometer arms, and the total working power differs by $\sim15$\% in two limits. Corresponding deviation of the Malus' law $\varepsilon(\theta)$ from periodic curve is clearly visible in the map $Z(\theta)$ in Figs.~\ref{fig3}({\bf a,b}), which may facilitate the difference between two particles since one is blocking illumination of the other.

The observations above suggest that the PP trap can be employed for simultaneous trapping of many particles. To elucidate this possibility we consider more complex multi-ring vortices created by the Laguerre-Gaussian beams LG$_{\rm ln}$, where the integer index $n$ indicates, in addition to the topological charge $l$,  the number of radial nodes (dark rings) in the transverse intensity distribution. Calculating the total intensity of the superposition of such co-rotating counter-propagating vortices we also assume a simple misalignment of their optical axes, namely a small relative tilt. This results in a complex light pattern with multiple minima shown in Fig.~\ref{fig4}({\bf a}). Experimental results are presented in Movies 5 and 6 with representative frames shown in Fig.~\ref{fig4}({\bf b,c}). In contrast to a single trap, in a multiple trap the particles strongly interact, as seen in Movie 5. Therefore, it remains unexplored whether the regular structures of many particles can be trapped in air. We estimate that the number of particles trapped in Fig.~\ref{fig4}({\bf a}) is around one hundred. Combining several such beams or employing holographic technique~\cite{tweezers08}, it is possible to create a `web' of vortex traps in a significant volume trapping a large number of particles.

\begin{figure}
\centering\includegraphics[width=8.5cm]{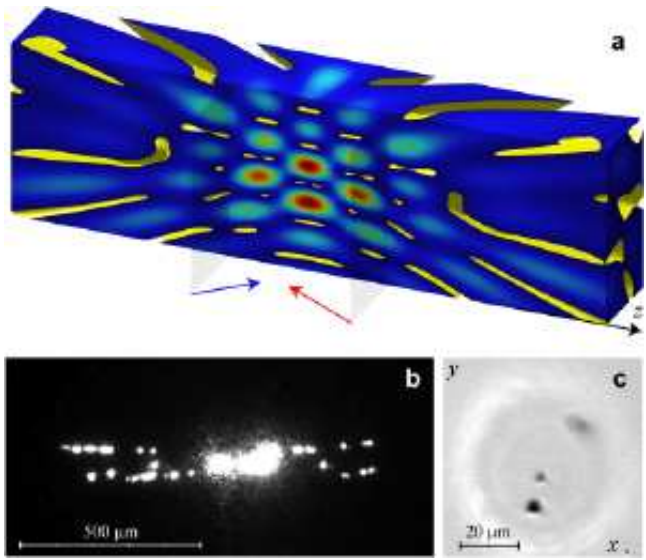}
\caption{{\bf Multiple PP trap with tilted beams.} {\bf a}, Volume plot of the longitudinal cut through the total intensity calculated for two counter-propagating Laguerre-Gaussian beams LG$_{12}$ tilted in the vertical direction by 0.02~rad. The yellow surfaces cut out the regions of small intensity where particles can be trapped. {\bf b}, The side view and {\bf c}, the front view of several particles simultaneously trapped with tilted beams.}
\label{fig4}
\end{figure}

{\bf SUMMARY.} 

We have demonstrated a novel all-optical method for trapping and manipulating small absorbing particles in gaseous media based on photophoretic force. We have realized experimentally, in open air, the robust three-dimensional guiding of agglomerates of carbon nanoparticles with the size spanned for two orders of magnitude, from 100~nm to $10\,\mu$m, over the distances of several millimeters, as well as their acceleration up to velocities of 1~cm/s and simultaneous trapping of a large number of particles. Our projections show that up-scaling of optical beam size will allow larger particles to be trapped and transported over longer distances, keeping trapping powers as low as few milli Watts. The ability of capture and controlled transport of nanoparticles in air by optical vortices may find wide applications in developing ecologically clean and health-safe environment for nanotechnology. The volume localization of nanoparticles will also allow engineering the appropriate control equipment for the manufacturing and micro-assembly processes. The nanoparticles agglomerated and collected in the non-contact and remote trap can be further investigated in terms of their chemical activity and unique toxicity (as compared to a bulk), important for health risk assessments. The outcomes are of fundamental importance for a wide range of other fields of science, such as interstellar dusty plasmas and atmospheric physics.\\


{\bf METHODS SUMMARY}

{\bf Photophoretic trap.} For the dual-vortex PP trap shown schematically in Fig.~\ref{fig1}({\bf a}), a particular care should be taken on the relative rotation, or topological charge, of two optical vortices. For paraxial beams the integer topological charge $l$ determines the order of a phase dislocation of the complex electric field~\cite{Berry,Soskin}, $E\sim\exp(il\varphi+ik_zz)$, where $k_z$ is the wave-number, $r$ and $\varphi$ are, respectively, the polar radius and angle in the transverse plane. Each time when the vortex beam is reflected, $k_z\to-k_z$, the sense of rotation remains unchanged, $l\to l$, so that vortex effectively reverses its topological charge~\cite{Shvedov}, defined with respect to the propagation direction $z$. Therefore, in the schemes based on the Fabry-Perot interferometers with even number of reflectors, the counter-propagating vortex beams, defined as $E_1\sim\exp(il_1\varphi+ik_zz)$ and $E_2\sim\exp(-il_2\varphi-ik_zz)$, will have opposite direction of rotation with the topological charges $l_1=l_2=l$, and the interference intensity pattern $|E_1+E_2|^2$ will be modulated azimuthally, i.e. $\sim \cos(2l\varphi+2k_z)$. The undesired intensity minima in the transverse intensity of the composite trapping beam will allow particles escaping, and it should be avoided. Therefore, we employ a scheme where a single vortex beam is reflected odd number of times before allowing to counter-propagate itself, similar to a shearing interferometer. The directions of rotation of the initial and reflected beam coincide in this case, $l_2=-l_1$, and azimuthal dependance $\exp(il_1\varphi)$ is factorized in the expression of the total field, thus the transverse intensity distribution remains radially symmetric. It is also noteworthy that such a constructive interference effectively doubles the optical angular momentum~\cite{Shvedov}, see the twisted power flow lines in Fig.~\ref{fig1}({\bf a}).\\

{\bf METHODS}

{\bf Synthesis of carbon nanoclusters.} To produce samples for this study, the graphite targets were ablated in a vacuum chamber pumped to a base pressure of $10^{-3}$~Torr and then filled with high-purity (99.995\%) argon gas. Carbon nanofoam, consisting of agglomerated nanoparticles of 4-6~nm in diameter, was synthesized using 40~W frequency doubled Nd:YVO$_4$ laser operating at wavelength $\lambda=532$~nm with repetition rate 1.5~MHz, pulse duration 12~ps~\cite{BLDAPA2004}, focused down to a spot size $\sim 15\,\mu$m, to generate laser intensity of $\sim 10^{12}$~W~cm$^{-2}$ with corresponding fluence of up to 20~J~cm$^{-2}$. The density of individual nanoparticles determined by the electron energy loss technique was in the range $(1.65\div 1.90)$~g~cm$^{-3}$~\cite{15,RodeAPA1999,RodeASS2002}, while the bulk density of the carbon nanofoam was varied in the range $(2\div 20)$~mg~cm$^{-3}$, depending on the argon pressure.

{\bf Thermal and optical properties of nanoclusters (carbon nanofoam).} Individual carbon nanoparticles have 80-90\% of $sp^2$ bonds~\cite{ArconPRB2006,LauPRB2007} and their ``bulk'' thermal and dielectric properties are similar to those of the graphite, namely the thermal conductivity is $k_g=6.3$~W~m$^{-1}$~K$^{-1}$~\cite{Pierce1993} and the refractive index $N_g=1.95+i0.66$ ($\varepsilon_g=N_g^2=3.4+i2.6$).

We assume that nanofoam consists of a mixture of graphite nanospheres and air with a volume filling fraction $\eta=(\rho_f-\rho_a)/(\rho_c-\rho_a)$. Taking the average density of the nanofoam $\rho_f = 10$~mg~cm$^{-3}$~\citet{15}, of air $\rho_a = 1.29$~mg~cm$^{-3}$, and of a nanoparticle $\rho_g = 1.8$~g~cm$^{-3}$, we obtain $\eta=4.84\times10^{-3}$. Due to very low density we find, using linearized Maxwell formula $k_f\simeq k_a(1+3\eta)$, that the thermal conductivity of the nanofoam, $k_f=0.0266$~W~m$^{-1}$~K$^{-1}$, is mainly determined by that of air, $k_a=0.0262$~W~m$^{-1}$~K$^{-1}$ at 300~K~\citet{HandbookC&P2008}.

Optical transmission measurements of the carbon nanofoam films with thicknesses varying in the range $70\div 120\,\mu$m yield the absorption length in the nanofoam $l_f= 35 \pm 5\,\mu$m (s.d.e.). Corresponding imaginary part $\kappa_f$ of the nanofoam refractive index, $N_f=n_f+i \kappa_f$, can be calculated as $\kappa_f=\lambda/4\pi l_f=(1.25\pm 0.15)\times10^{-3}$ (s.d.e.). On the other hand, the dielectric function of low density absorbing mixture can be found using the effective medium approximation, namely the linearized Maxwell Garnet formula~\cite{LL1984}: $\varepsilon_f= \varepsilon_a+3\eta\varepsilon_a (\varepsilon_g-\varepsilon_a)/(\varepsilon_g+2\varepsilon_a)$. Taking the dielectric constant of air $\varepsilon_a=1$, we obtain for nanofoam $\varepsilon_f=1.0079+i0.0032$ and $N_f=1.0040+i0.0016$. The value of imaginary part of refractive index is in good agreement with that derived from experimental data.

The reflectivity of the surface of the nanofoam films, and thus of the nanoclusters, is very low. The measured diffused radiation shows no angular dependence characteristic to Lambertian source; the fraction of the incident radiation power scattered into $4\pi$ steradian was $(2.84\pm 0.14)\times10^{-2}$ (s.d.e.). The power fraction of the reflected light at normal incidence was as low as $(1.8 \pm 0.3)\times10^{-6}$ (s.d.e.), similar to the value reported recently for the arrays of carbon nanotubes~\cite{NanL_08}.

{\bf Evaluation of optical forces.} Theoretical treatment of the PP force was developed for spherical particle illuminated by a plane wave. Therefore, we consider a sphere with typical radius $a=1\,\mu$m and with thermal and optical parameters of the nanofoam, illuminated by a plane wave with characteristic intensity $I_0=P/4\pi w^2=1.1$~kW~cm$^{-2}$, here the vortex ring radius $w_0= 8.4\,\mu$m and the typical power $P= 0.01$~W. Corresponding radiation pressure force, $F_a=P_a/c = 4.5\times 10^{-15}$~N, is one order of magnitude larger than the gravitational force, $F_g=4.1\times 10^{-16}$~N, here the power absorbed by the particle $P_a= Pa^3/3w_0^2l_f=1.35\times 10^{-7}$~W. The absorbtion efficiency of the particle (ratio of the absorbed to incoming power) is of the order of a few percent, $\xi=4a/3l_f=0.038$.

Evaluation of the PP force is essentially more difficult. It involves solving consistently the electrodynamic and gas-kinetic problems taking into account thermophysical, optical, and accommodation properties of an aerosol particle~\cite{book2002,BeresnevPhFl1993}. We apply the limit of total accommodation and low Knudsen number, $Kn=l/a=0.065$, here $l=65$~nm is the mean-free path of air molecules. In this regime the PP force is a result of the Maxwellian ``thermal creep" of the gas molecules along the temperature gradient on the particle surface and the gas is modeled as continuous fluid media with boundary slip-flow conditions. The expression for the PP force is given by (see, e.g., formula (34) in ref.~\cite{BeresnevPhFl1993})
\[F_{\rm pp}=-J_1 \frac{9\pi\mu_a^2 a I_0}{2\rho_a T (k_f+2k_a)},\]
here the air viscosity $\mu_a=1.73\times10^{-5}$~N~s~m$^{-2}$ at temperature $T=298$~K. The most important parameter which defines the sign and magnitude of the PP force is the asymmetry factor, $-0.5\le J_1\le 0.5$, characterizing the angular nonuniformity of the heat sources at the particle surface. Depending on whether the front or back surface of the particle is hotter the particle will move away (positive PP and $J_1<0$) or toward the light source (negative PP and $J_1>0$). Negative PP was observed for semi-transparent particles focusing illumination on the back side while for black body particles $J_1=-0.5$ and PP is positive. Although the nanofoam skin-depth is larger than typical nanocluster dimension, the thermal conductivity is very low, thus we assume that the temperature changes on the particle surface are determined by the illuminating intensity, similar to the black body particles. The factor $J_1$, however, takes into account corresponding absorption efficiency, similar to $\xi$ above, thus we adopt the following value: $J_1=-\xi/2=-2a/3l_f$. In these assumptions the PP force dominates radiation pressure by four orders of magnitude, $F_{pp}=3\times10^{-11}$~N.

However, the forces exerted by a light with spatially varying intensity, such as that of a vortex beam, can be essentially different. We estimate in Supplementary Notes I that the actual radiation pressure force $F_a$ in our experiments is two orders of magnitude lower, $F_a=10^{-16}$~N, because the particle is located on the dark vortex axis and absorbs much less power, $P_a =P 8a^5/15w^4l_f=3.1\times10^{-8}$~W. Thus the radiation pressure of a ``hollow'' vortex beam alone might be insufficient to counterbalance gravity. Similar reduction by a factor $2a^2/w^2$ is predicted for the PP force which became $F_z=8.4\times 10^{-13}$~N, at least three orders of magnitude larger than gravitation.

Finally, the radial trapping of particles by the transverse PP force $F_R$, exerted by the vortex ring, is possible because there is no reduction of the force magnitude, it scales as $F_R/F_z=-8R/3a$, here $R$ is the radial displacement of the particle from on-axis stationary position, see Supplementary Notes II. In contrast, the transverse trapping by the radiation pressure (gradient force) requires strong focusing of the vortex beam by a microscope objective with high numerical aperture, NA$\ge 1$, and the radial force is one order of magnitude lower than the axial (longitudinal) force~\cite{AQC_98_469}.

\vspace{5mm}
\begin{center}{\large \bf Supplementary information}\end{center}




\section*{Supplementary Methods: Experimental setup}

\begin{figure*}
\centering\includegraphics[width=18cm]{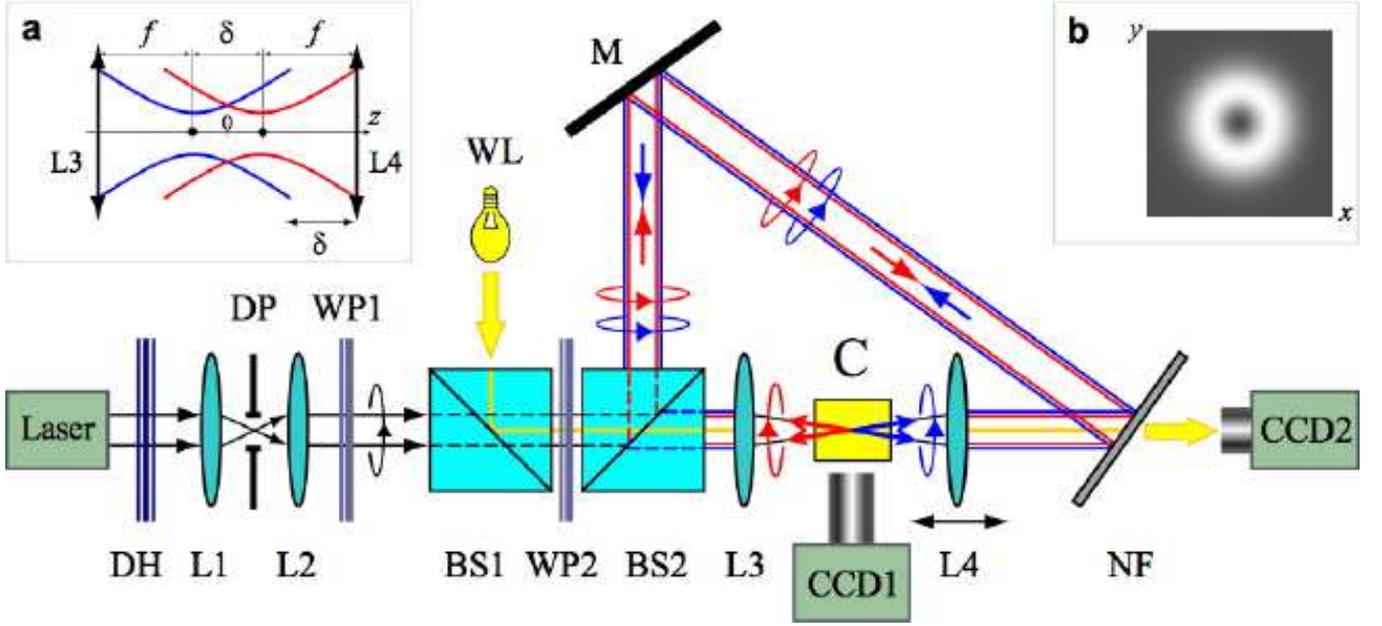}
\caption{{\bf Experimental setup of an optical trap for absorbing particles created by the standing wave of two counter-propagating vortex beams.} {\bf a}, Dual-beam trap with movable lens L4 adjusting the separation of focal planes $\delta$. {\bf b}, The ring-like transverse intensity distribution of a Laguerre-Gauss vortex beam. Setup elements: DH -- diffraction hologram, L -- lenses, DP -- diaphragm, WP -- half-wave plates, BS -- polarizing beam-splitters, WL -- white light source, M -- mirror, C -- glass cell, NF -- notch filter.}
\label{fig1}
\end{figure*}

Our experimental setup is drawn in Fig.~5. The linearly polarized Gaussian beam from a cw laser source with variable power (Verdi V5, Coherent Inc., wavelength $\lambda=532$nm) passes through the diffraction ``fork''-type hologram DH where it is transformed into a Laguerre-Gauss vortex beam with topological charge $l=1$ and transverse intensity pattern shown in the inset ({\bf 5b}). The beam diameter can be varied from 2.5mm to 8mm by a collimator based on two lenses, L1 and L2. The half-wave plate WP1 adjusts polarization of the input vortex beam so that it passes through the polarizing beam splitter BS1; the later serves as an injector of white light from the source WL to monitor the transverse dynamics of particles trapped in the glass cell C. The white light provides the background illumination for the images of the particles at the CCD2 camera after passing the notch filter NF which cuts off the laser radiation.

The interferometer consists of tree reflectors -- a notch filter NF reflecting the laser beam and passing the background light through, a mirror M, and a polarizing beam splitter cube BS2. The beam splitter BS2 divides the vortex beam into two beams: the {\em forward}-propagating beam (blue arrows) and the {\em backward}-propagating beam (red arrows). The forward beam passes through the lenses L3 and L4, reflects from the notch filter NF and mirror M, and exits trough BS2. The backward beam from BS2, after refection from M and NF, enters the system in the opposite direction and goes through L4, L3, to exit the interferometer reflecting from BS2. The scheme is designed so that both beams have only a single round trip, thus preventing unwanted interference of the beams on the second path. The interferometer is formed with the odd number of reflectors so that the axial symmetry of the intensity distribution is preserved for any polarization state of counter-propagating beams, see the discussion in Methods Summary. A particle trapping volume C is formed between the lenses L3 and L4 and the distance $\delta$ between their focal planes can be varied by moving lens L4, see the inset ({\bf 5a}). The imaging camera CCD1 collects the light scattered by the particles and monitors the behavior of the trapped particles in the longitudinal cross-section.

The half-wave plate WP2 allows gradually change the tilt angle $\theta$ of the polarization of the input optical vortex and thus to control the ratio $\varepsilon=P_f/P_b$ of the powers of forward $P_f$ and backward $P_b$ beams after the beam splitter BS2 with low extinction ratio 1:13. For full characterization of this important parameter we measured the powers of both beams inside the interferometer for two orthogonally polarized states, i.e., for $\theta=0$ and $\theta=\pi/4$. Applying Malus' law we derive the expression
\be
\varepsilon(\theta)=\frac{1}{\gamma}\frac{\alpha \cos^2 2\theta+\beta \sin^2 2\theta}{1-\alpha \cos^22\theta-\beta \sin^22\theta},\;\;\; 0.093\le\varepsilon\le15.623,
\label{SE1}
\ee
here $\alpha=0.928$ and $\beta=0.071$ are the coefficients of transmission through BS2 for two orthogonal linear polarizations and $\gamma=0.825$ is the transmission coefficient through the long arm of the interferometer taking into account loses of the backward beam on the mirror M and notch filter NF. The total working power $P=P_f+P_b$ inside the interferometer is less than the power $P_{\rm in}$ passing from the laser onto the beam-splitter cube BS1 because of loses $\gamma<1$. It can be calculated as $P/P_{\rm in}=\gamma+(1-\gamma) (\alpha\cos^22\theta+\beta\sin^22\theta)$ and it varies slightly with $\theta$, $0.837\le P/P_{\rm in}\le 0.987$.

\section*{Supplementary Notes}

In addition to the estimation of the magnitudes of optical forces in Methods section, based on the theory developed for the spherical particles illuminated by a plane wave, here we attempt to calculate forces induced by a ring-shaped vortex beam. In particular, we show that the fraction of optical vortex power absorbed by a particle is two orders of magnitude lower than that from a plane wave and, although the full treatment of photophoresis problem is beyond the scope of this work, we introduce a simple model which allows straightforward calculation of the trapping position in good correspondence with experimental data.

\subsection*{Supplementary Notes I: Calculation of the longitudinal forces}

{\bf Radiation pressure force.} The estimation of the radiation pressure force $F_a=P_a/c$ exerted on a particle requires calculation of the absorbed power $P_a$. For a better comparison with the PP theory for a plane wave illumination we first introduce the characteristic intensity of such hypothetical plane wave, $I_0=P/4\pi w^2$ (see Fig.~\ref{fig2}({\bf a})), where $P=\int I d{\vec \rho}$ is the beam power of an optical vortex with the intensity
\be
I(\rho,z)=\frac{P}{\pi}\frac{\rho^2}{w^4(z)}\exp\left(-\frac{\rho^2}{w^4(z)}\right).
\label{I}
\ee
Here the ring radius $w(z)=w_0\sqrt{1+z^2/z_0^2}$ with the beam waist $w_0$, the diffraction length $z_0=2\pi w_0^2/\lambda$, and $\rho$ is a polar radius in the plane transverse to the optical axis $z$. To calculate the absorbed power $P_a$ we use the Beer–-Lambert law for the intensity $I_t$ transmitted through the nanofoam film of the thickness $l_z$, $I_t=I_{\rm in}\exp(-l_z/l_f)$, here $I_{\rm in}$ is the incident intensity and $l_f=35\,\mu$m, see Methods. The power $P_a=\int (I_{\rm in}-I_t)d\vec{\rho}$ and for a spherical particle with the radius $a$ we have $l_z=2\sqrt{a^2-\rho^2}$ and $P_a=2\pi\int_0^a I_{\rm in} \left\{1-\exp\left(-2\sqrt{a^2-\rho^2}/l_f\right)\right\}\rho d\rho$. For a plane wave $I_{\rm in}=I_0$ and
\be
F_a= \frac{P_a}{c}=\frac{\pi a^2I_0}{c}\left\{1-\frac{l_f^2}{2a^2}\; f\left(\frac{2a}{l_f}\right)\right\}, 
\label{Pa}
\ee
where  the function  $ f(t)=1-(1+t)e^{-t}$.
It is important for the following to note that, for small argument $t\ll1$, the function $f(t)\simeq t^2 (1/2-t/3)$. For the representative particle with $a=1\,\mu$m ($2a/l_f=0.057$) we approximate the absorbed power as $P_a\simeq 4\pi I_0 a^3/3l_f$; in terms of the vortex power $P$ corresponding force can be expressed as
\[F_a\simeq \frac{P}{c} \frac{a^3}{3w^2l_f}, \text{ for a plane wave } I_{\rm in}=I_0.\]
Taking typical values $P=0.01$~W and $w\ge w_0=8.4\,\mu$m we obtain the radiation pressure force $F_a=4.5\times 10^{-15}$~N, one order of magnitude larger than the gravitational force, $F_g=mg=4.1\times 10^{-16}$~N, here $g=9.81$~m~s$^{-2}$ is the standard gravity and the particle mass $m=4\pi\rho_f a^3/3=4.2\times10^{-14}$~g. Note that, for small particles $a\ll l_f$, the ratio $F_a/F_g=P/4\pi w^2 l_f \rho_f g$ does not depend on the particle radius $a$. Other parameters are $I_0=1.1$~kW~cm$^{-2}$, $P_a=1.35\times 10^{-6}$~W, and the absorption efficiency $\xi=P_a/P_{\rm in}=4a/3l_f=0.038$, here the incoming power $P_{\rm in}=\pi a^2 I_0=3.54\times 10^{-5}$~W.

For a spatially varying intensity $I_{\rm in}=I(\rho,z)$ we assume that the spherical particle with radius $a$ is located on optical axis, as shown in Fig.~2~({\bf a}), at the distance $z=Z$ from the beam waist at $z=0$. In our experiments the ring radius $w\ge w_0=8.4\,\mu$m and $z_0=837.3\,\mu$m, thus we can neglect the variation of the vortex intensity on the distances $Z-a \le z\le Z$, comparable with the particle size $a=1\,\mu{\rm m}\lesssim w_0 \ll z_0$, taking
\begin{eqnarray}
w^2\left(z-Z\right) &=& w_0^2\left(1+\frac{(z-Z)^2}{z_0^2}\right)\simeq \\ \nonumber & & \simeq w_0^2\left(1+\frac{Z^2}{z_0^2}\right)=w^2\left(Z\right).
\label{w}
\end{eqnarray}
The incoming power is given by $P_{\rm in}=P\,f\left(a^2/w^2\right)$ and, for particles smaller than the vortex ring $a\ll w$, we use the expansion of the function $f(t)$ after \ref{Pa} to obtain $P_{\rm in}\simeq Pa^4/2w^4 = 10^{-6}$~W. It is a factor of $2a^2/w^2\le 0.028$ smaller than the corresponding power of a ``plane-wave'' above, $P_{\rm in}=Pa^2/4w^2$, because of the intensity zero at the vortex origin, see the red-shaded region in Fig.~2~({\bf a}). Similarly, the absorbed power is two orders of magnitude lower, $P_a=3.1\times10^{-8}$~W, than that for a plane wave above (the absorption efficiency is of course the same $\xi=16a/15l_f=0.031$); to the leading order of the small parameter $a/w\ll1$ the radiation pressure force is given by
\[F_a \simeq \frac{P}{c} \frac{8a^5}{15w^4l_f}, \text{ for a vortex beam } I_{\rm in}=I(\rho,Z).\]
Therefore, the force $F_a= 10^{-16}$~N exerted by a vortex beam is insufficient to compensate gravity for a chosen level of optical power. Note, however, that the ratio $F_a/F_g$ scales as $a^2$ (limited by the assumptions above $a\ll w$ and $a\ll l_f$), so that the radiation pressure can overcome gravity for larger particles and for higher optical power.\\

\begin{figure}
\centering\includegraphics[width=8cm]{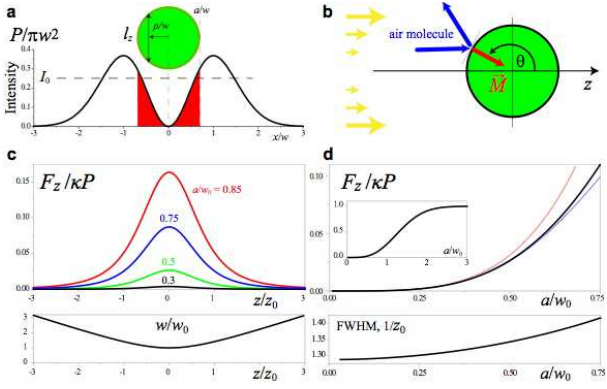}
\caption{{\bf To the calculation of the longitudinal forces.} {\bf a}, Red-shaded is the part of a vortex beam illuminating particle (green sphere). {\bf b}, Transfer of a momentum (red arrow) from a gas molecule to a particle; the illuminated side of the particle is a hemisphere $\pi\le\theta\le\pi/2$. {\bf c}, The PP force \ref{Fz} versus axial distance $z$, the bottom frame shows the ring radius $w(z)$ of the diffracting vortex beam. {\bf d}, Amplitude ($f(a^2/w_0^2)$, top) and the full width at half maximum (FWHM, bottom) of the PP force curves shown in ({\bf c}). The red and blue curves in ({\bf d}) show the lowest order approximations of \ref{Fz}: $f(t)\simeq t^2/2$ (red) as in \ref{FZ}, and taking into account next term $f(t)\simeq t^2(1/2-t/3)$ (blue).}
\label{fig2}
\end{figure}

{\bf Photophoretic force.} As we briefly discussed in Methods, the calculation of the PP force involves solving consistently two problems, the electrodynamic radiation problem including absorption and re-emission, and the gas-kinetic equations taking into account thermophysical and accommodation properties of an aerosol particle. Most of the works on PP employ Mie scattering theory to solve the first problem, therefore, the available solutions, such as Eq.~(1) in Methods, were derived for a spherical particle illuminated by a plane wave. Here we attempt to overcome this limitation and take into account the spatially varying vortex intensity. In general, the solution to the electrodynamic problem with non-uniform illumination can be obtained, for spherical particles with given optical properties, as a superposition of known Mie solutions for plane waves. However, the nanoclusters have random shape and they are composite of carbon nanoparticles, thus the optical properties are determined by a complex multiple scattering and can be only estimated using rough effective medium approach, see Methods. Therefore, instead of involving calculations with Mie solutions we adopt a somewhat intuitive theoretical approach.

As we demonstrated above the amount of optical power absorbed by a nanocluster is very low because of a large absorption depth in nanofoam, therefore, we can neglect the distortion of vortex beam intensity by the particle. Then, because of low thermal conductivity we assume that the temperature changes on the particle surface are determined by the illuminating intensity $I$, similar to the black body particles. Finally, we introduce the linear momentum flux, $|\vec{M}|=\kappa I$, transferred from gas molecules to the particle, as shown in Fig.~\ref{fig4}({\bf b}). The phenomenological coefficient $\kappa$ has the dimension of inverse velocity and absorbs thermal and optical parameters of the gas and the particle. By definition, the PP force is given as an integral over the momentum flux density, $\vec{F}=\int \vec{M} dS$, here $dS=2\pi a^2\sin\theta d\theta$ is particle surface element taking into account cylindrical symmetry of the problem. The longitudinal PP force $F_z$ is given by (see Fig.~\ref{fig4}({\bf b}))
\be
F_z=\int_{S_+} M_z dS,\;\;\; M_z=-\kappa \cos\theta\, I(S_+),
\label{Fz}
\ee
here $S_+$ is illuminated hemisphere, $\pi/2\le\theta\le\pi$. For illumination by a plane wave with characteristic intensity $I_0$ we obtain $F_z=\kappa \pi a^2 I_0$. Comparison with the expression for PP force, $F_{pp}=3\times10^{-11}$~N, derived in Methods, allows us to evaluate
\[\kappa =-J_1\frac{9\mu_a^2}{2 a \rho_a T(k_f+2 k_a)}\simeq\]
\[\simeq \frac{3 \mu_a^2}{l_f \rho_a T (k_f+2 k_a)}=8.5\times 10^{-7}\;\frac{\rm s}{\rm m},\]
here the last expression is obtained with $J_1=-\xi/2=-2a/3l_f$ (see Methods), derived above for small particles $a\ll l_f$, the thermal conductivity of the nanofoam $k_f=0.0266$~W~m$^{-1}$~K$^{-1}$ and the parameters of air are: viscosity $\mu_a=1.73\times10^{-5}$~N~s~m$^{-2}$, temperature $T=298$~K, mass density $\rho_a = 1.29$~mg~cm$^{-3}$, and thermal conductivity $k_a=0.0262$~W~m$^{-1}$~K$^{-1}$. Note that the final expression for $\kappa$ does not depend on particle radius and optical power, it characterizes the transfer of momentum from air molecules to the material with given properties.

For a vortex beam the intensity $I(S_+)$ in \ref{Fz} is taken from \ref{I} with the approximation \ref{w}. Integration in \ref{Fz} gives
\[
F_z=\kappa P \; f\left(\frac{a^2}{w^2}\right),
\]
with the function $f(t)$ defined in \ref{Pa}. For small particles $a\ll w$ the expression can be greatly simplified,
\be
F_z\simeq \kappa \frac{P}{2}\frac{a^4}{w^4},
\label{FZ}
\ee
and it differs from the solution for plane wave illumination above, $F_z=\kappa P a^2/4w^2$, by a factor of $2a^2/w^4\le 0.028$, exactly the same reduction as for radiation pressure force. Therefore, the actual force exerted by a vortex beam on our representative spherical nanocluster with $a=1\,\mu$m is $F_z=8.4\times10^{-13}$~N.

The PP force \ref{Fz} can be measured in units of $\kappa P$ and it is visualized in Fig.~\ref{fig2}({\bf c}, {\bf d}). The bell-shaped curves in ({\bf c}) appear because of vortex diffraction, $w(z)$, shown in the bottom panel for better comparison. As seen in ({\bf d}) the amplitude of the force saturates for large particles $a>2w_0$ so that, for the particles with $a\lesssim0.5 w_0$, the approximation \ref{FZ} is sufficiently accurate while for larger particles the next term in the expansion of $f(t)$ can be taken to improve the results. The full width at half maximum (FWHM) of the curves in ({\bf c}) has weak dependance on the size parameter $a/w_0$ as seen in the bottom panel in ({\bf d}).


%

\subsection*{Supplementary Notes II: Calculation of transverse PP force}

\begin{figure}
\centering\includegraphics[width=8cm]{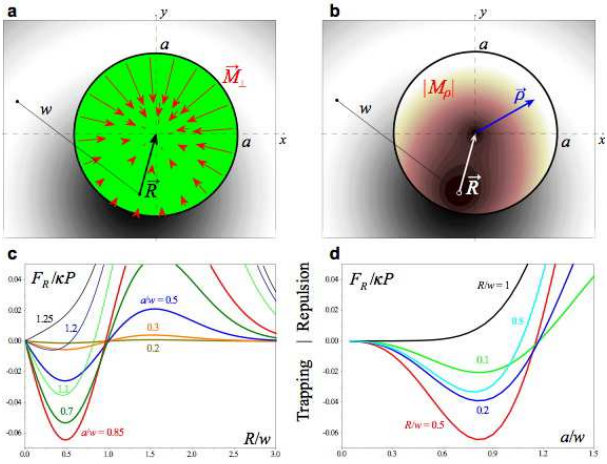}
\caption{{\bf To the calculation of transverse PP force.} {\bf a}, Geometry of the problem; arrows show the transverse projection of the momentum flux density $\vec{M}_\perp$ transferred to the particle (green sphere); corresponding amplitude $|\vec{M}_\perp|=|M_\rho|$ is color-coded in {\bf b}. The background in pictures ({\bf a}, {\bf b}) is the grey-coded intensity of a vortex with the ring radius $w$. {\bf c}, {\bf d}, The magnitude of the transverse force~\ref{Fy} versus two key parameters, $R/w$ and $a/w$.}
\label{fig4}
\end{figure}

To estimate the geometry and the magnitude of PP force in transverse plane $\vec{F}_\perp$ and evaluate its trapping efficiency we adopt the same assumptions as for the calculation of the longitudinal force above, namely we assume a spherical strongly absorbing particle with the radius $a$ much smaller than the diffraction length $z_0$ so that \ref{w} applies. Furthermore, from the geometry of the momentum transfer depicted in Fig.~\ref{fig2}({\bf b}) we can see that the modulus of the area density of the transverse momentum flux is given by the projection (cf.~\ref{Fz})
\[|\vec{M}_\perp|=\kappa \sin\theta\, I(S_+),\]
while its direction is towards the particle origin. Since $\sin\theta = \rho/a$, where $\rho$ is the polar radius in the transverse cross-section of the particle, we derive:

\[\vec{M}_\perp = -\kappa \frac{\vec{\rho}}{a}\, I(S_+),\]
\[\vec{F}_\perp \equiv \int_{S_+} \vec{M}_\perp dS = - \kappa \,\int_{S_+} \frac{\vec{\rho}}{a}\, I(S) dS.\]  

We assume that the particle is shifted by a vector $\vec{R}$ from the optical axis, then the distribution of the momentum flux density is mirror-symmetric with respect to $\vec{R}$ as shown in Fig.~\ref{fig4}({\bf a}). There are two zeros in the distribution of the amplitude $|\vec{M}_\perp|$ as seen in ({\bf b}), one zero at the origin of the vortex intensity ring and another at the origin of a particle because of zero projection at $\theta=\pi$. The combination of two minima produce highly asymmetric distribution along $\vec{R}$ which is the source of the PP force returning particle to the vortex origin. From the mirror symmetry follows that the transverse force is antiparallel $\vec{R}$ and directed towards vortex axis, $\vec{F}_\perp=\frac{\vec{R}}{R}F_R$ and $F_R\le 0$. Therefore, for simplicity and without loss of generality we can choose, e.g., the cartesian projections $R_x=0$ and $R_y=R$,
\begin{eqnarray}
F_R &=&  -\frac{\kappa P}{\pi} \iint_{x^2+y^2\le a^2}\, \frac{y}{a}\, \frac{x^2+(y+R)^2}{w^4(Z)}\times \\ \nonumber
 & & \times \exp\left(\frac{-x^2-(y+R)^2}{w^2(Z)}\right) \,\frac{adxdy}{\sqrt{a^2-x^2-y^2}}.
\label{Fy}
\end{eqnarray}
Transverse force is calculated as a function of two key parameters, $R/w$ and $a/w$, the results are presented in Fig.~\ref{fig4}({\bf c}, {\bf d}). Several conclusions can be readily made. First, there is a sharp boundary between the trapping region inside the beam ($R<w$) and the repulsion of particles far from optical axis ($R>w$); the unstable equilibrium ($F_\rho=0$) is located approximately at the ring radius $R \lesssim w$. Second, while the small particles are always trapped in the vicinity of optical axis, for particles larger than the ring $a\gtrsim w$ the trapping efficiency rapidly vanishes; the maximal trapping force is achieved for particles with $a\simeq0.85 w$. Finally, the remarkable scaling property of~\ref{Fy} leads to a somewhat counterintuitive result: the transverse force on a large particle ($a>0.85 w$) can be stronger for lower vortex intensity when particle moves away from the beam focus. Indeed, the radius $w(Z)$ increases because of the diffraction and the parameter $a/w$ decreases, thus the particle effectively follows the curves in ({\bf d}) from right to left passing by the force maximum. The reason for this unexpected result, as seen in ({\bf a, b}), is the interplay of geometrical contributions to the momentum flux versus spatial structure of the vortex intensity ``weighting'' momentum flux density in~\ref{Fy}.

Transverse force vanishes at the vortex axis, $R\to 0$, as should be expected from symmetry considerations. Close to this equilibrium position and for small particles, $a+R\ll w_0$, we can linearize~\ref{Fy} and obtain
\be
-F_R=\kappa P\frac{4}{3}\frac{Ra^3}{w^4(Z)}\le \kappa P\frac{4}{3}\frac{Ra^3}{w_0^4}.
\label{Fys}
\ee
In our horizontal scheme the transverse PP force should be compared to the gravitational force, $F_g=4\pi \rho_f\, a^3 g/3$, here $g=9.81$~m~s$^{-2}$ is the standard gravity and $\rho_f=10$~mg~cm$^{-3}$ is the characteristic mass density of the nanofoam, see Methods. Because the PP force~\ref{Fys} scales as $a^3$, similar to $F_g$, it is possible to increase particle size and still balance gravitation without increasing the laser power $P$.

\subsection*{Supplementary Notes III: Calculation of the trap position}

Based on the setup in Fig.~\ref{fig1} we have two principal degrees of freedom for spatial manipulation of trapped particles, apart from the total power $P$ of the beams which defines the magnitude of the PP force. One is the distance $\delta$ between focal spots of two counter-propagating beams, adjusted by the position of the lens L4. Another is the tilt $\theta$ of the half-wave plate WP2 defining the power ratio of two beams $\varepsilon(\theta)$ in \ref{SE1}. Each beam applies a PP force given by \ref{Fz} being a function of the position of the particle, here $z$, relative to the beam focal plane $z=0$. For two beams with focuses separated by the distance $\delta$ we introduce the origin of the coordinates in the middle between focal planes located at $z=\pm\delta/2$, see Fig.~\ref{fig1}({\bf a}). The total longitudinal force on the particle is then given by the sum, $F_{\rm tot}=F_z\left(z-\delta/2;P_f\right)-F_z\left(z+\delta/s;P_b\right)$, so that using \ref{Fz} we obtain
\begin{eqnarray}
\frac{F_{\rm tot}(z)}{\kappa}&=&P_f\,f\left(\frac{a^2/w_0^2}{1+\left(z+\delta/2\right)^2/z_0^2}\right) \\ \nonumber
&-& P_b\,f\left(\frac{a^2/w_0^2}{1+\left(z-\delta/2\right)^2/z_0^2}\right),
\label{Ft}
\end{eqnarray}
here $P_f$ and $P_b$ are the powers of the forward and backward beams, respectively. Stationary points $z=Z$ can be found from the condition $F_{\rm tot}(Z)=0$. As we demonstrated above, for the particles with $2a < w_0\le w(z)$ the linearized expression for the force \ref{FZ} provides good approximation. In this limit the stationary point (trap position $Z$) does not depend on the particle radius $a$ and it is given by the solutions to a corresponding quadratic equation
\be
Z^\pm=\frac{\delta}{2}\frac{\epsilon+1}{\epsilon-1} \pm \sqrt{\frac{\delta^2\epsilon}{(\epsilon-1)^2}-z_0^2},
\label{Zpm}
\ee
here $\epsilon^2(\theta)=\varepsilon(\theta)$ from \ref{SE1}. However, the stationary point can be stable or unstable. Linearizing \ref{Ft} in the vicinity of a stationary point, $z=Z+\xi$ with $|\xi|\ll |Z|$, we obtain $F_{\rm tot}(Z+\xi)\simeq\xi\;\left( dF_{\rm tot}/dz\right)|_{z=Z}$. Therefore, the stationary point $Z$ is {\em stable} if $\left( dF_{\rm tot}/dz\right)|_{z=Z}<0$ (the force is returning), and it is {\em unstable} otherwise. Calculation of the derivative of the total force \ref{Ft} gives the following result:
\be
\left. \left( \frac{dF_{\rm tot}}{dz}\right)\right|_{z=Z^\pm}= \pm \,\delta\, Z^\pm\,C, \text{   here   } C\left(\delta, Z^\pm\right)>0.
\label{stab}
\ee
Therefore, there is one stable and one unstable stationary point, see Fig.~\ref{fig3}({\bf a}), and the stability is defined by the sign of the root in \ref{Zpm} as well as the sign of the inter-focal distance $\delta$. For equal powers of two beams, $P_f=P_b$ and $\epsilon=1$ at $\theta=\pi(0.134, 0.366, 0.634, 0.866)$, the force derivative is proportional to $-\delta$, instead of \ref{stab}, and the trapping position $Z=0$ is stable only for $\delta>0$, when focal planes of two beams are separated as shown in Fig.~\ref{fig1}({\bf a}). The case $\delta<0$ corresponds to the anti-separated focuses; although there are stable trapping positions, they are located outside the inter-focal region where the transverse force rapidly decreases, see Supplementary Notes II above. Furthermore, the trapping is possible (the roots in \ref{Zpm} are real) only if the separation of focuses $\delta$ exceeds some minimal value, $|\delta|\ge z_0|\epsilon-1|/\sqrt{\epsilon}$. In the opposite case the real roots disappear and there is no stationary trapping on axis. Similarly, if we use an ideal polarizing beam splitter BS2 in our scheme, so that $\alpha=1$ and $\beta=0$ in \ref{SE1}, the power ratio would become unbounded $\epsilon=\tan2\theta/\sqrt{\gamma}$ and, for any $\delta$, the trap position with disappear (move to infinity $Z\to \infty$) for some value of $\theta$. It  could still be used for stationary guiding (positioning) of particles, as in the manuscript figure~3({\bf a-c}), but the dynamical bouncing of particles, as in the manuscript figure~3({\bf d-f}), will be impossible. In contrast, for our ``unperfect'' beam-splitter BS2 we have bounded power ratio, $0.304\le\epsilon(\theta)\le 3.953$, and there is the stationary solution \ref{Zpm} for any $0\le\theta\le 2\pi$ (full turn of the input polarizer) if $\delta \ge 1.485 z_0$.

\begin{figure}
\centering\includegraphics[width=8cm]{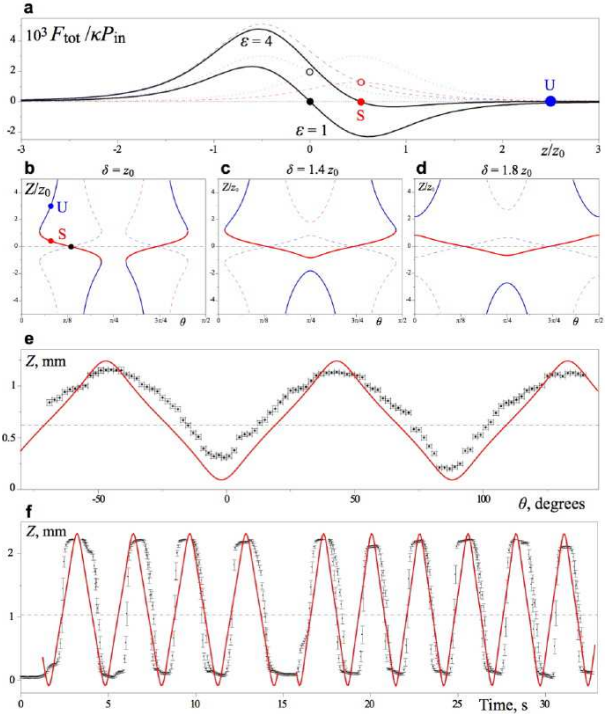}
\caption{{\bf To the calculation of the trap position.} {\bf a}, The total longitudinal forces (black solid lines) for equal ($\varepsilon=1$) and unequal ($\varepsilon=4$) powers at the inter-focal distance $\delta=z_0$. Corresponding stable (S) and unstable (U) positions of the trap are indicated in the $Z(\theta)$ diagram in ({\bf b}). Similar diagrams ({\bf c}) and ({\bf d}) show stable (red) and unstable (blue) roots $Z(\theta)$ for different values of $\delta$: solid lines for $\delta>0$ and dashed lines for $\delta<0$. The calculated position of the trap (red lines) in ({\bf e}) and ({\bf f}) is compared to experimental data (black bars) from the manuscript figures 3({\bf a}) and 3({\bf d}), respectively.}
\label{fig3}
\end{figure}

Our objective here is to find the domain $L$ of stable trapping along optical axis as a function of $\delta$ for the given characteristic of a beam-splitter \ref{SE1}. For separated focal planes $\delta>0$ this domain includes a middle point $Z=0$ for equal powers and it is given by $L= Z^-(\epsilon_{\max})-Z^+(\epsilon_{\min})$, here

\medskip\noindent
{\it (i) the inter-focal distance $\delta < 1.261\, z_0$}: stable trapping is limited by a finite domain of $\theta$, for a given $\delta$ the limits satisfy $\epsilon_{\max,\min}=1+\Delta/2 \pm \Delta\sqrt{1+\Delta^2/4}$, here $\Delta=\delta/z_0$, see an example in Fig.~\ref{fig3}({\bf b}). The domain is $L=\sqrt{\delta^2+4z_0^2}$ so that, for $\delta \to 0$ and $\epsilon_{\max,\min}\to 1\pm \delta/z_0$, we formally obtain the minimal value of the domain $L \to 2z_0$. However, at the limit $\delta=0$ and $\epsilon=1$ the force is zero everywhere, so that close to this limit the magnitude of force is infinitesimally small. The domain growths with $\delta\to 1.261\, z_0$ and $L_{\max}=2.364\, z_0$.

\medskip\noindent
{\it (ii) the inter-focal distance $1.261<\delta /z_0 < 1.485$}: here $\epsilon_{\min}=0.304$ and $\epsilon_{\max}=1+\Delta^2/2 + \Delta\sqrt{1+\Delta^2/4}$. The domain $L$ decreases monotonically from $2.364\, z_0$ to $2.015\, z_0$, the trapping is continuous on a limited interval of $\theta$, see Fig.~\ref{fig3}({\bf c}).

\medskip\noindent
{\it (iii) the inter-focal distance $1.485\, z_0<\delta$}: trapping is continuous for $\theta\in [0,2\pi)$ with $\epsilon_{\min}=0.304$ and $\epsilon_{\max}=3.953$, see Fig.~\ref{fig3}({\bf d}). The trapping domain, $L/z_0= 1.7762 \Delta-\sqrt{0.4534 \Delta^2-1}-\sqrt{0.6290 \Delta^2-1}$, first decreases to reach its minimum $L_{\min}=1.3737\,z_0$ at $\delta=2.4351\,z_0$, then diverges with increasing $\delta$.\\

We compare theoretical calculations with experimental results in Fig.~\ref{fig3}({\bf e,f}). Both experiments were done for continuous trapping in the whole domain $\theta\in[0,2\pi)$, described in the case {\em (iii)} above with the corresponding roots shown in Fig.~\ref{fig3}({\bf d}). For the static guiding (positioning) in Fig.~\ref{fig3}({\bf e}), as in the manuscript figure 3({\bf a}), the inter-focal distance $\delta = 2.04$~mm corresponds to the minimal value of the trapping domain $L=1.15$~mm, slightly larger than the observed guiding over the distance $\sim 1$~mm. However, the slopes of the curve in Fig.~\ref{fig3}({\bf e}) agree well with experimental data points. It is not the case in Fig.~\ref{fig3}({\bf f}), where the results of {\em dynamical} guiding, from the manuscript figure 3({\bf d}), are compared to the calculations of {\em stationary} trap position. Indeed, the slopes of the experimental dataset are much steeper than that of the solid theoretical curve because the slope of the curve $Z(t)$ corresponds to the particle velocity, the latter reaches values up to $\sim 1$~cm/s, determined by measuring the longest particle tracks (vertical bars in ({\bf f})) $\sim 400\,\mu$m recorded with exposition 40~ms. Furthermore, the extrema of the experimental dataset are flatter than that of the theoretical curve, this difference is a direct manifestation of the inertia of trapped particles moving along the optical axis and spending more time close to the points of return. The discontinuity visible in the middle of the dataset in ({\bf f}) is because the rotation of the half-wave plate was not monotonic (simply by hand). More importantly, the theoretical predictions on the trapping domain $L\simeq 2.4$~mm, calculated for the inter-focal distance $\delta\simeq 7.3$~mm used in experiment, agree well with the limits of particle motion $L_{\rm exp}\simeq 2.2$~mm. We conclude that relatively simple theoretical model developed here can be used to estimate and predict the domain of the PP trapping.


\begin{thebibliography}{99}

\bibitem{tweezers08} Dholakia, K., Reece, P. \& Gu, M. Optical micromanipulation. {\em Chem. Soc. Rev.} {\bf 37}, 42-55 (2008).

\bibitem{AshPRL70} Ashkin, A. Acceleration and trapping of particles by radiation pressure. {\em \prl} {\bf 24}, 156-159 (1970).

\bibitem{AshOL86} Ashkin, A., Dziedzic, J. M., Bjorkholm, J. E., \& Chu, S. Observation of a single-beam gradient force optical trap for dielectric particles. {\em \ol} {\bf 11}, 288-290 (1986).

\bibitem{book2002} Davis, E. J. \& Schweiger, G. {\em The Airborne Microparticle: Its Physics, Chemistry, Optics, and Transport Phenomena.} (Springer, 2002), pp. 780-785.

\bibitem{AshSCI80} Ashkin, A. Applications of Laser Radiation Pressure. {\em Science} {\bf 210}, 1081-1088 (1980).

\bibitem{PZ_17_352} Ehrenhaft, F. On the physics of millionths of centimeters. {\em Phys. Z.} {\bf 18}, 352-368 (1917).


\bibitem{Preining_66} Preining, O. Photophoresis. {\em Aerosol Sciences} Ed. C. N. Davies (Academic Press, New York, 1966), pp. 111-135.


\bibitem{14} Gamaly, E. G. \& Rode, A. V. Nanostructures Created by Lasers. {\em Encyclopedia of Nanoscience and Nanotechnology} {\bf 7}, 783-809 (American Scientific Publishers, Stevenson Range, 2004).

\bibitem{15} Rode, A. V., Gamaly, E. G. \& Luther-Davies, B. Formation of cluster-assembled carbon nano-foam by high-repetition-rate laser ablation. {\em Appl. Phys. A} {\bf 70}, 135-144 (2000).

\bibitem{Berry} Nye, J.~F. \& Berry, M.~V. Dislocations in wave trains. {\em Proc. R. Soc. London A} {\bf 336}, 165-190 (1974).

\bibitem{Soskin} Soskin, M. S. \& Vasnetsov, M. V. Singular Optics. {\em Prog. Opt.} {\bf 42}, 219-276 (Ed. E.~Wolf, Elsevier, Amsterdam, 2001).

\bibitem{EPJE_04_287} Steinbach, J., Blum, J. \& Krause, M. Development of an optical trap for microparticle clouds in dilute gases. {\em Eur. Phys. J. E} {\bf 15}, 287-291 (2004).


\bibitem{OberdorPatFibTox2005} Oberd\"orster, G. {\em et al.} 
    Principles for characterizing the potential human health effects from exposure to nanomaterials: elements of a screening strategy. {\em Particle and Fibre Toxicology} \textbf{2}:8 (2005).

\bibitem{MaynardNat2006} Maynard, A. D. {\em et al.} 
    Safe handling of nanotechnology. {\em Nature} {\bf 444}, 267-269 (2006).

\bibitem{NelSci2006} Nel, A., Xia, T., M\"adler, L. \& Li, N. Toxic Potential of Materials at the Nanolevel. {\em Science} {\bf 311}, 622-627 (2006).

\bibitem{ColvinNatBio2003} Colvin, V. L. The Potential Environmental Impact of Engineered Nanomaterials. {\em Nat. Biotechnol.} {\bf 21}, 1166-1170 (2003).



\bibitem{JGR_67_455} Hidy, G. M. \& Broc, J. R. Photophoresis and the descent of particles into the lower stratosphere {\em J. Geophys. Res.} {\bf 72}, 455-460 (1967).


\bibitem{PRL_06_134301} Wurm, G. \& Krauss, O. Dust Eruptions by Photophoresis and Solid State Greenhouse Effects. {\em \prl} {\bf 96}, 134301-4 (2006).

\bibitem{AA_07_977} Krauss, O. {\em et al.} 
    The photophoretic sweeping of dust in transient protoplanetary disks. {\em Astron. \& Astrophys.} {\bf 462}, 977-987 (2007).

\bibitem{AA_07_L9} Mousis, O. {\em et al.} 
    Photophoresis as a source of hot minerals in comets. {\em Astron. \& Astrophys.} {\bf 466}, L9-L12 (2007).


\bibitem{AQC_98_469} Rubinsztein-Dunlop, H., Nieminen, T. A., Friese, M. E. J. \& Heckenberg, N. R. Optical trapping of absorbing particles. {\em Adv. Quant. Chemistry} {\bf 30}, 469-492 (1998).

\bibitem{tweezers03} Grier, D. G. A revolution in optical manipulation. {\em Nature} {\bf 424}, 810-816 (2003).

\bibitem{bio} Svoboda, K. \& Block, S. M. Biological applications of optical forces. {\em Annu. Rev. Biophys. Biomol. Struct.} {\bf 23}, 247-285 (1994).

\bibitem{molecule} Neuman, K. C., Lionnet, T. \& Allemand, J.-F. Single-Molecule Micromanipulation Techniques. {\em Annu. Rev. Mater. Res.} {\bf 37,} 33-67 (2007).

\bibitem{chu} Chu, S. The manipulation of neutral particles. {\em Nobel Lectures, Physics 1996-2000} (Ed. G. Ekspong, World Sc. Pub. Co., Singapore, 2002), pp. 122-158.

\bibitem{air} McGloin, D. {\em et al.} 
    Optical manipulation of airborne particles: techniques and applications. {\em Faraday Discuss.} {\bf 137}, 335-350 (2008).

\bibitem{APL_82_455} Lewittes, M., Arnold, S. \& Oster, G. Radiometric levitation of micron sized spheres. {\em Appl. Phys. Lett.} {\bf 40}, 455-457 (1982).

\bibitem{BeresnevPhFl1993} Beresnev, S., Chernyak, V. \& Fomyagin, G. Photophoresis of a spherical particle in rarefied gas. {\em Phys. Fluids A} \textbf{5}, 2043-2052 (1993).

\bibitem{JCS_64_50} Rosen, M. H. \& Orr, C. The photophoretic force. {\em J. Colloid Sci.} {\bf 19}, 50-60 (1964). 

\bibitem{Shvedov} Alexeyev, C. N., Yavorsky, M. A. \& Shvedov, V. G. Angular momentum flux of counter-propagating paraxial beams. {\em \josa B} {\bf 25}, 643-646 (2008).

\bibitem{BLDAPA2004} Luther-Davies, B. {\em et al.} 
    Table-Top 50 W Laser System for Ultra-Fast Laser Ablation. {\em Appl. Phys. A} {\bf 79}, 1051-1055 (2004).

\bibitem{RodeASS2002} Rode, A. V. {\em et al.} 
    Electronic and magnetic properties of carbon nanofoam produced by high-repetition-rate laser ablation, {\em Appl. Surf. Sci.} {\bf 197-198}, 644-649 (2002).

\bibitem{RodeAPA1999}Rode, A. V. {\em et al.} 
    Structural analysis of a carbon foam formed by high pulse-rate laser ablation. {\em Appl. Phys. A} {\bf 69}, S755-S758 (1999).

\bibitem{ArconPRB2006}Ar\v{c}on, D. {\em et al.} 
    Origin of Magnetic Moments in Carbon Nanofoam. {\em Phys. Rev. B} {\bf 74}, 014438 (2006).

\bibitem{LauPRB2007} Lau, D. W. M., McCulloch, D. G., Marks, N. A., Madsen, N. R. \& Rode, A. V. High-Temperature Formation of Carbon Onions within Nanofoam: An Experimental and Simulation Study. {\em Phys. Rev. B} {\bf 75}, 233408 (2007).

\bibitem{LL1984} Landau, L. D., Lifshitz, E. M. \& Pitaevskii, L. P. {\em Electrodynamics of Continous Media.} (Pergamon Press, Oxford, 1984), p.44.

\bibitem{Pierce1993} Pierce, H. O. {\em Handbook of carbon, graphite, diamond and fullerenes} (Park Ridge, NJ, Noyes Publications, 1993).

\bibitem{HandbookC&P2008} {\em CRC Handbook of Chemistry and Physics}, Ed. Lide, D. R., 88th ed. (CRC, Taylor \& Francis Group, Boca Raton Fl, 2008).




\bibitem{NanL_08} Yang, Z. P., Ci, L., Bur, J. A., Lin, S. Y. \& Ajayan, P. M. Experimental Observation of an Extremely Dark Material Made By a Low-Density Nanotube Array. {\em Nano Lett.} {\bf 8}, 446-451 (2008). 

\end{thebibliography}
\end{document}